\documentclass[]{spie}  

\newcommand{\bs}{\boldsymbol}
\usepackage{amsmath,amsfonts,amssymb}
\usepackage{graphicx}
\usepackage[colorlinks=true, allcolors=blue]{hyperref}

\title{A generic FPGA-based detector readout and real-time image processing board
}

\author{Mayuresh Sarpotdar, Joice Mathew, Margarita Safonova and Jayant Murthy}

\affil{Indian Institute of Astrophysics, Bangalore, India}

\authorinfo{Further author information: (Send correspondence to Mayuresh Sarpotdar)\\Mayuresh: E-mail: mayuresh@iiap.res.in\\  Joice: E-mail: joice@iiap.res.in}

\begin{document} 
\maketitle

\begin{abstract}
For space-based astronomical observations, it is important to have a mechanism to capture the digital output from the standard detector for further on-board analysis and storage. We have developed a generic (application-wise) field-programmable gate array (FPGA) board to interface with an image sensor, a method to generate the clocks required to read the image data from the sensor, and a real-time image processor system (on-chip) which can be used for various image processing tasks. The FPGA board is applied as the image processor board in the Lunar Ultraviolet Cosmic Imager (LUCI) and a star sensor (\emph{StarSense}) -- instruments developed by our group. In this paper, we discuss the various design considerations for this board and its applications in the future balloon and possible space flights.  
\end{abstract}


\section{INTRODUCTION}
\label{sec:intro}  

Data generated in astronomical payloads in space or high-altitude balloon is large in size. To download this data through the limited downlink capability available on a satellite or a high-altitude balloon, the on-board processing of the data is required to reduce its size. We have developed an FPGA-based generic board which can be used for this task, and this utility can be implemented in various digital interfaces without much change in the hardware.  

For any image processing board, some of the basic required components are a controller for image acquisition from the detector, RAM for temporary storage of the images, a microprocessor for processing the image data, and a permanent memory for on-board storage. Our FPGA board combines all of these requirements. This board can be used to read the digital output from any standard semiconductor detector, e.g. CMOS or CCD. The acquired image is sent to RAM, and is processed by the microprocessor embedded in the FPGA. The processed image is stored in a long-term memory (SD card), and can be either transmitted to ground station from space, or retrieved in case of a balloon flight. In Section~\ref{sec:board_design} we describe the steps followed to design the board. A brief description of the image sensors that can be used in conjunction with this board is given in Section~\ref{sec:sensors}, and a brief description of algorithms for different image processing tasks to pinpoint the processing capabilities of the board is presented in Section~\ref{sec:processing}. The conclusions and future applications of this processing board are given in the Section~\ref{sec:conclusion}.

\section{FPGA BOARD DESIGN}
\label{sec:board_design}

\subsection{Components of the FPGA board}

This board was developed as a generic application with intentions to use it as an image processor board in the flight instruments developed by our group --- Lunar Ultraviolet Cosmic Imager (LUCI)[\citenum{{luci1},{luci2}}] and a star sensor {\it StarSense}[\citenum{{starsense1},{starsense2}}]. The board is to be interfaced with the ultraviolet (UV) CCD on LUCI, and with a radiation-hardened CMOS sensor Star1000 on {\it StarSense}. The characteristics and readout sequences of these sensors are described later in Section~3.

\begin{figure} [h!]
\begin{center}
\begin{tabular}{c} 
\includegraphics[height=7cm]{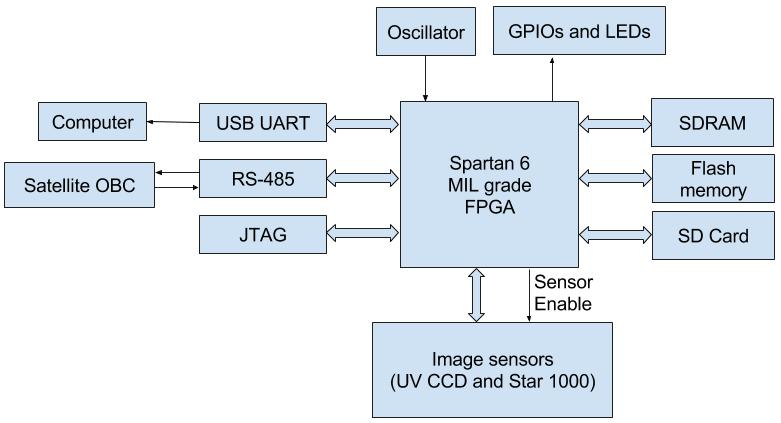}
\end{tabular}
\end{center}
\caption[]{ \label{fig:block_dia} Block diagram of the FPGA board}
\end{figure}

The main components of the FPGA board are shown in Table~\ref{tab:tech_specs} and their layout is shown in Fig.~\ref{fig:block_dia}. An  FPGA is an integrated circuit designed to be configured by a programmer, or end user, to carry out a specific set of tasks. An FPGA contains programmable logic components called logic blocks, and a hierarchy of re-configurable interconnects that allow the blocks to be wired together to form different configurations. We have selected a MIL-grade Spartan-6Q FPGA from Xilinx\footnote{http://www.xilinx.com} because of its wide operating temperature range. The program for the FPGA is written in a hardware description language called Verilog. We also implement a soft core microprocessor\footnote{http://www.xilinx.com/products/design\_resources/proc\_central/microblaze\_faq.pdf}, called microblaze, inside the FPGA. The program for microblaze is written in C. The Verilog code is synthesized and a configuration program is generated using the programming toolsuite ISE from Xilinx. The configuration program for the FPGA is known as a bitstream. The FPGA consists of configurable logic blocks where the configuration is volatile, that is the FPGA has to be reprogrammed at each power cycle. We have selected a Master Serial programming interface for the FPGA, wherein the FPGA reads its bitstream from a flash memory on its own using a serial transfer protocol, called serial peripheral interface (SPI), and configures itself. The size of the bitstream determines the size of the flash memory chip. In our case, for a Spartan-6Q FPGA device, the bitstream size is 4 MB. Other than the bitstream, the flash memory can be used to store data that should be permanently available to the FPGA e.g. a star catalog, some configuration parameters for the payload, etc. Therefore we have selected a flash memory chip of 8 MB. The image sensors which we use typically have a size of $1024 \times 1024$ pixels, each pixel with a 10-bit digital value. This corresponds to an image size of $\sim 1.2$ MB being processed by the FPGA every 100 ms. We have selected a 64 MB RAM, which is sufficient for image processing tasks on images from the selected detectors. For the permanent storage of processed data, we use an SD card of 32 GB.  The FPGA  board can be connected to the on-board computer (OBC) of a satellite, or a balloon payload, through the standard RS485 protocol. During the development cycle, the FPGA can be connected to a computer through a USB port. 

\begin{table}[h!]
\caption{Technical Specifications of the FPGA board} 
\label{tab:tech_specs}
\begin{center}       
\begin{tabular}{|l|l|} 
\hline
\rule[-1ex]{0pt}{3.5ex} {\bf Component} & {\bf Specifications} \\
\hline
\rule[-1ex]{0pt}{3.5ex}  FPGA & Spartan-6Q (XQ6SLX150T-2FGG484)\\
\hline
\rule[-1ex]{0pt}{3.5ex}  Size (mm)& $65\times 65$ \\
\hline
\rule[-1ex]{0pt}{3.5ex}  Weight & 50 gms \\
\hline
\rule[-1ex]{0pt}{3.5ex}  Power  & 3 W \\
\hline
\rule[-1ex]{0pt}{3.5ex}  RAM$^\dagger$   & 64 MB SDRAM (MT46V32M16) \\
\hline
\rule[-1ex]{0pt}{3.5ex} Flash memory$^\dagger$  & 8 MB chip (N25Q064A)\\
\hline
\rule[-1ex]{0pt}{3.5ex}  SD card  & 32 GB \\
\hline
\rule[-1ex]{0pt}{3.5ex}  Connectivity to the On-Board Computer  & RS485 (MAX481), USB-to-UART (FT232)\\
\hline
\end{tabular}
\end{center}
\hskip 0.7in $^\dagger$ Micron Technology, Inc., USA. {\tt https://www.micron.com/}
\end{table}

\subsection{Voltage regulators}

\begin{figure} [h!]
\begin{center}
\begin{tabular}{c} 
\includegraphics[height=4cm]{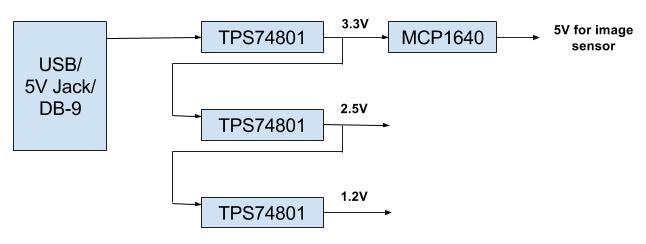}
\end{tabular}
\end{center}
\caption[]{ \label{fig:power_supply} Power supply for the FPGA board}
\end{figure}

The FPGA board can be powered by an external computer (a laptop, a satellite/balloon OBC) through either a USB cable or through a DB-9 (9 pins) connector (Fig.~\ref{fig:block_dia}). In addition to this, it can also be powered by a DC jack provided on the board. The FPGA needs 3 different voltages: 3.3 V, 2.5 V and 1.2 V. It needs 3.3 V for general-purpose input/output pins, image sensor interface, SD card, flash memory interface, and RS485 interface. Voltage of 2.5 V is needed to connect to the SDRAM. Finally, it needs 1.2 V for its internal functioning. All these voltages are generated from the unregulated 5 V input through the voltage regulators. The regulated voltage output should be switched on in a sequence to ensure successful programming of the FPGA. The FPGA is set in master serial configuration mode, and therefore, the flash memory chip should be powered on and ready to accept memory read commands, before the FPGA starts sending them. Therefore, the voltage regulators are connected in a chain as shown in Fig.~\ref{fig:power_supply}. The programmable soft-start feature of the voltage regulator ensures that the flash memory is powered on at least 3 ms prior to the FPGA, and is ready to accept read commands from the FPGA. The image sensors require a 5 V supply, which can be switched on or off in order to reduce power consumption, when not in use. We do not use the 5 V input directly, instead a boost regulator is used to increase the voltage from 3.3 to 5 V. Such configuration gives a more stable voltage for the image sensors, which are the critical part of the circuit.  

\section{IMAGE SENSORS}
\label{sec:sensors}

The image sensors in LUCI and \emph{StarSense} are directly connected with the FPGA board. LUCI uses a CCD sensor (ICX407BLA) from Sony, which is sensitive in ultraviolet wavelengths, and \emph{Starsense} uses a radiation-hardened CMOS sensor (NOIS1SM1000A) from ON Semiconductors. Details of these image sensors are given in the following subsection.

\subsection{Ultraviolet CCD}

The UV CCD is a diagonal 8 mm (type 1/2-inch) interline CCD solid-state image sensor with a square pixel array and 1.45 Megapixels (Mp). Its technical details are shown in Table~\ref{tab:image_sensor_specs}. This chip features an electronic shutter with variable charge-storage time which makes it possible to realize full-frame still images. The effective quantum efficiency of this detector in near-UV wavelengths (200 to 400 nm) is approximately 20\%.  

\begin{figure} [ht]
\begin{center}
\begin{tabular}{ccc} 
\includegraphics[height=4cm]{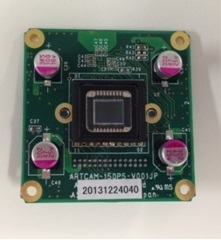} &
\includegraphics[height=4cm]{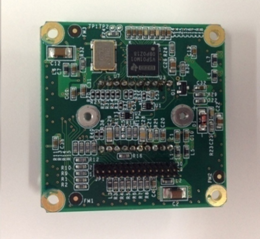} &
\includegraphics[trim=350 1200 350 1200, clip, height=4cm]{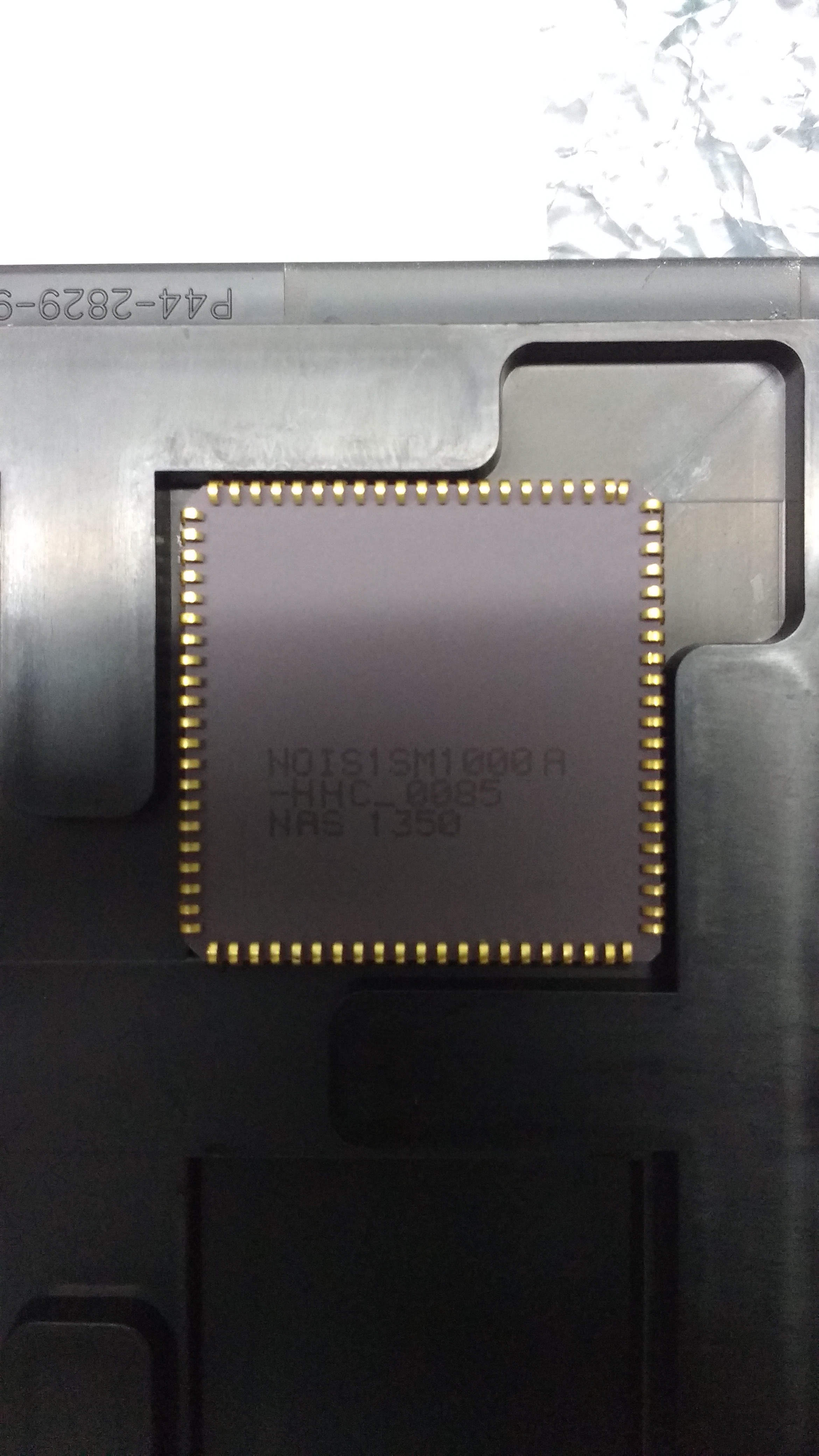}
\end{tabular}
\end{center}
\caption{{\it Left, Middle}: Photos of the UV CCD PCB, front and back side, respectively. {\it Right}: A photo of the STAR1000 image sensor chip.}
\label{fig:uvccd_photo}
\end{figure}

This camera has a stack of 3 PCBs. First PCB contains the UV CCD and the timing generator chip, which is required to generate reset clocks and transfer clocks required for the CCD (Fig.~\ref{fig:uvccd_photo}, {\it Left} and {\it Middle}.). Second PCB contains the voltage regulators to provide various voltages for reset gate, vertical transfer and horizontal transfer clocks. Third PCB contains its own FPGA which transmits the digital data received from the CCD PCB to the computer through USB. This PCB does not store the image data on-board or process it. We have used the CCD PCB and the voltage regulator PCB as is and connected our FPGA board to them. 

\begin{table}[h]
\caption{Specifications of the UV CCD and Star1000} 
\label{tab:image_sensor_specs}
\begin{center} 
\begin{tabular}{l ll} 
 \hline
\rule[-1ex]{0pt}{3.5ex}{\bf Sensor} &  ICX407BLA UV CCD & NOIS1SM1000A Star1000  \\\hline
\rule[-1ex]{0pt}{3.5ex}{\bf Array size (mm)} &  $6.47\times 4.83$ &  $15\times 15$ \\
\rule[-1ex]{0pt}{3.5ex}{\bf Total number of pixels} &  $1360\times 1024$ &  $1024\times 1024$\\
\rule[-1ex]{0pt}{3.5ex}{\bf Pixel size ($\bs{\mu}$m) }& $4.65\times 4.65$ & $15\times 15$\\
\rule[-1ex]{0pt}{3.5ex}{\bf Frame rate (fps)} & 12 & 11 \\
\rule[-1ex]{0pt}{3.5ex}{\bf Voltage} & 5 V & 5 V\\
\rule[-1ex]{0pt}{3.5ex}{\bf Power consumption} & 2 W & 0.4 W  \\
\hline
\end{tabular}
\end{center}
\end{table} 

\subsection{Star1000}

A star sensor is a critical sensor in satellite or balloon missions. The image sensor is the only electronics part which is open to incident radiation. Degradation of the detector quality with exposure to intense radiation cannot be tolerated and therefore, a radiation-hardened CMOS image sensor STAR1000\footnote{ON Semiconductors, USA. {\tt http://www.onsemi.com}.}  was selected for this application  (Fig.~\ref{fig:uvccd_photo}, {\it Right}). Its technical details are shown in Table~\ref{tab:image_sensor_specs}. It can also support the readout of small windows of a whole image 
at much faster frame rates. The sensor features the on-chip fixed-pattern noise (FPN) correction, a programmable gain amplifier, and a 10-bit analog-to-digital converter (ADC). The registers, which contain $X-$ and $Y-$addresses of the read out pixels, can be directly accessed by an external controller. This architecture provides flexible operation and allows different operation modes, such as (multiple) windowing, sub-sampling, and so on.

\section{READOUT PROCESS}
\label{sec:readout_process}

\begin{figure} [ht]
\begin{center}
\begin{tabular}{c} 
\includegraphics[trim=0 100 0 100,clip, height=6cm]{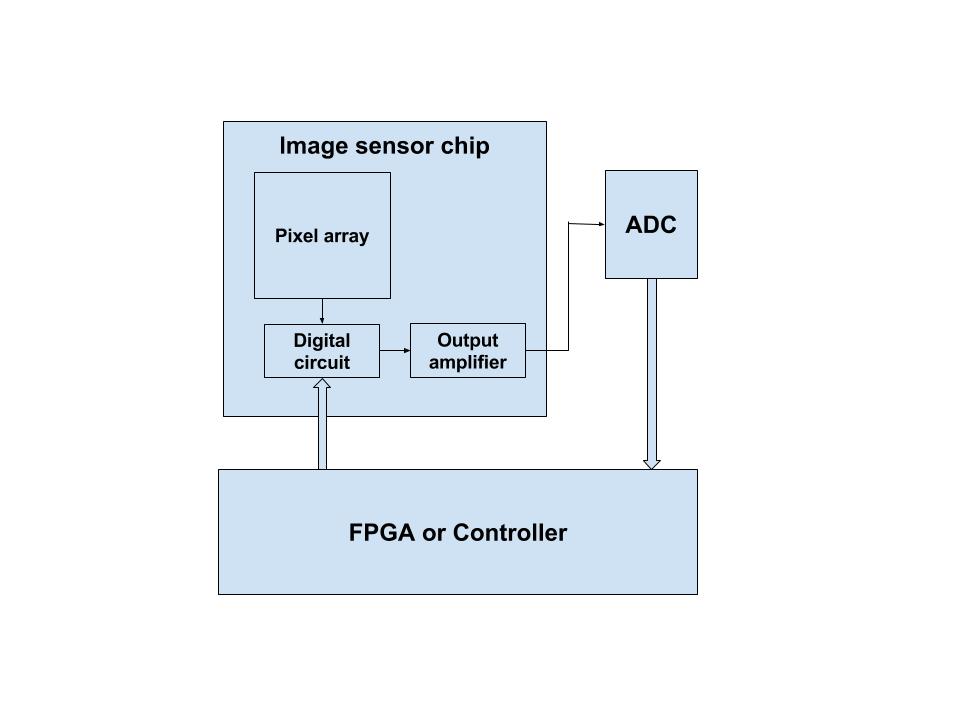}
\end{tabular}
\end{center}
\caption[]{ \label{fig:readout_process} General schematics of a readout process.}
\end{figure}

An image sensor is essentially an array of photodiodes (pixels) which measure the light intensity at each pixel. The output of each pixel is an analog voltage. Each pixel is connected to an output amplifier on the chip through digital circuitry. The clock signals control this digital circuitry and manage the sequence of connections. The analog voltage of the output amplifier is converted to a digital value by an ADC. The external FPGA, or a controller, generates all the clock signals to control the digital circuit and ADC, and captures the digital output from the ADC. Figure~\ref{fig:readout_process} gives a general outline of this process. The process of collecting digital data (a measure of the light intensity on each pixel) is called the readout process. In this section, we describe how the readout process is implemented for both sensors. 

\subsection{UV CCD}

\begin{figure}[h!]
\begin{center}
\begin{tabular}{c} 
\includegraphics[trim = 80 400 100 125, clip, height=4cm]{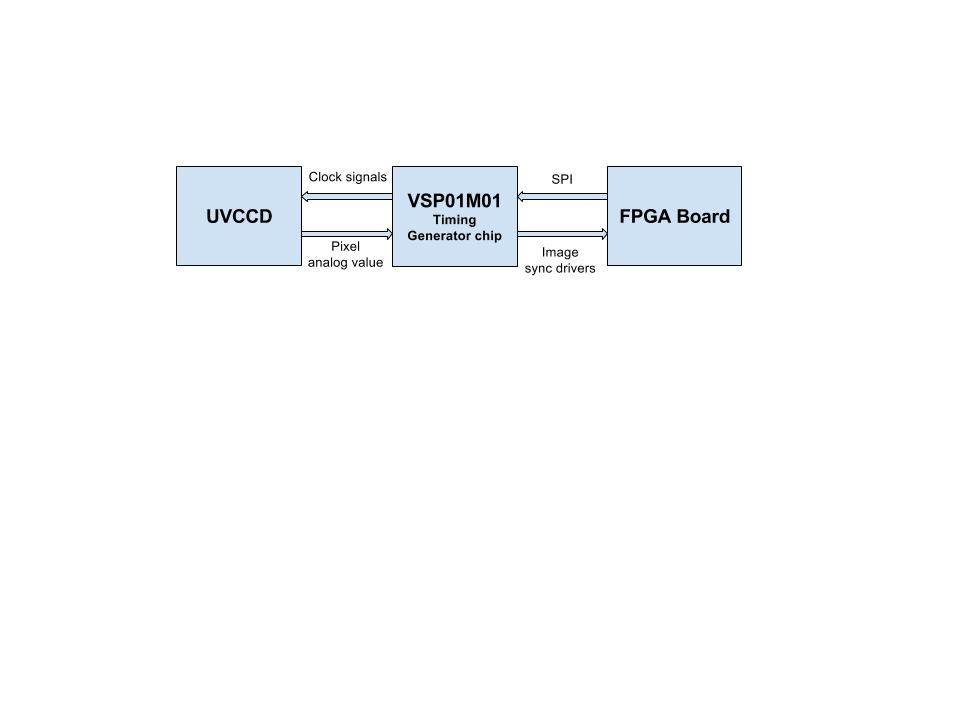}
\end{tabular}
\end{center}
\caption[]{ \label{fig:uvccd_readout} A layout of the UV CCD readout circuit.}
\end{figure}

As mentioned in Section~\ref{sec:sensors}.1, there is a timing generator chip VSP01M01\footnote{Texas Instruments, USA. {\tt www.ti.com}} along with the image sensor on the same PCB. 
The required timing for various clock signals are to be programmed on the chip when the circuit is switched on. This programming is done by the microblaze, implemented on the FPGA board using SPI protocol. The timing generator chip includes the digital circuitry to access the analog voltage from each pixel and an ADC. It also generates image synchronization driver signals (frame-valid, line-valid, and pixel clock) which are received by the FPGA board. The FPGA board captures digital value of each pixel and decodes the position of the pixel in the image using these signals. Figure~\ref{fig:uvccd_readout} shows a layout of UV CCD readout circuit.

\subsection{Star 1000}

\begin{figure}[h]
\begin{center}
\begin{tabular}{c} 
\includegraphics[trim = 0 30 0 0, clip, height=10cm]{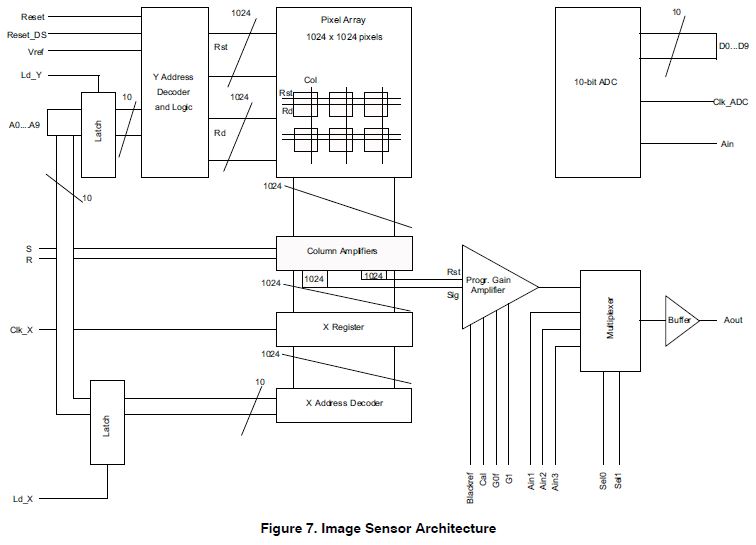}
\end{tabular}
\end{center}
\caption[]{ \label{fig:star1000_arch} Internal architecture of Star1000 image sensor. Source: ON Semiconductors datasheet[\citenum{datasheet}].}
\end{figure} 

The Star1000 image sensor does not have a timing generator chip like the UV CCD. Therefore, we have to generate the clock signals for various pins on the image sensor from the FPGA using Verilog code. The image sensor consists of 5 basic modules as shown in Fig.~\ref{fig:star1000_arch}:
\begin{enumerate}
\item Pixel array. 
\item $X$--$Y$-addressing logic. 
\item Column amplifier. 
\item Output amplifier.
\item Analog to Digital Converter (ADC).
\end{enumerate}
The pixel array is the light-sensitive part of the image sensor. It consists of $1024 \times 1024$ pixels each of size $15\,\mu$m$\times 15\,\mu$m (Table~\ref{tab:image_sensor_specs}). The pixel to be read out is selected using the address value given on the address bus and a proper signal to the $Ld\verb|_|X$ and $Ld\verb|_|Y$ pins (Fig.~\ref{fig:star1000_arch}). The $X-$ and $Y-$ addressing logic decodes the location of the pixel from the address bus. The column amplifier samples the output voltage and the reset level of the pixel whose row is selected, and presents these voltage levels to the output amplifier. The output amplifier combines subtraction of pixel signal level from reset level with a programmable gain amplifier. Finally the ADC converts the analog output from the output amplifier to a digital value. It converts the analog value to a 10-bit digital value which can be sampled by the FPGA. The FPGA board is programmed to generate clock signals to control these modules. These clock signals control the image acquisition from the image sensor. 

\begin{figure}[h]
\begin{center}
\begin{tabular}{c} 
\includegraphics[scale=0.5]{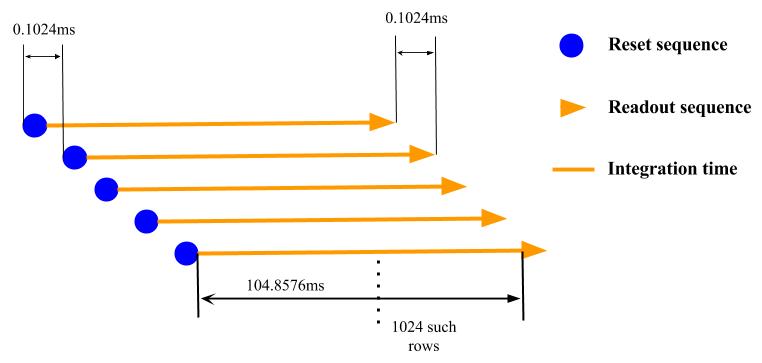}
\end{tabular}
\end{center}
\caption{\label{fig:star1000_readout_overall} General procedure to read out images from Star1000.}
\end{figure}

The image acquisition process is implemented into two separate processes: reset and readout. The reset process can be done row-wise and the readout process can be done pixel-wise. The readout process is done in two parts: row readout and pixel readout.  The clock signals for reset process, row readout process, and pixel readout process are described below. A general sequence for the readout of the image sensor is shown in Fig.~\ref{fig:star1000_readout_overall}.

\begin{figure}[h!]
\begin{center}
\begin{tabular}{c} 
\includegraphics[height=11cm]{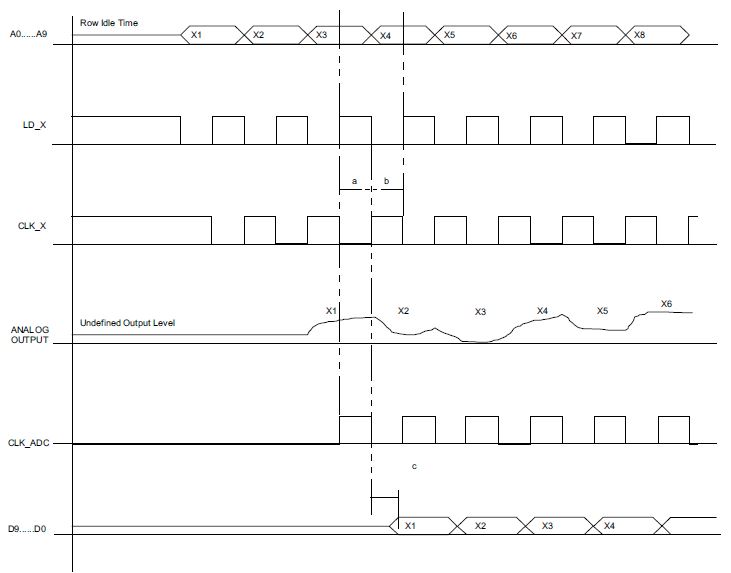}
\end{tabular}
\end{center}
\caption[]{ \label{fig:star1000_pixel_read} Timing diagram for column-wise pixel readout process[\citenum{datasheet}].}
\end{figure}

\begin{itemize}

\item{Row reset sequence: 
A row in the image sensor can be reset by the following sequence (illustrated in Fig.~\ref{fig:star1000_reset_seq}):
\begin{enumerate}
\item{Place proper $Y-$address on address pins.}
\item{Assert $Ld\verb|_|Y$ and latch the address into the internal decoder circuitry.}
\item{Pulse the `Reset' pin so that the internal decoder circuitry resets the $Y-$row in the image sensor.}
\end{enumerate}
}

\begin{figure}[h!]
\begin{center}
\begin{tabular}{c} 
\includegraphics[height=4cm]{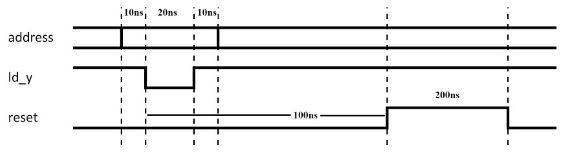}
\end{tabular}
\end{center}
\caption[]{ \label{fig:star1000_reset_seq} Timing diagram for row reset process.}
\end{figure} 

\item{Row readout sequence:
After reset process is done and after the integration time is elapsed, each row must be read. By this process, the outputs of the pixels in the row are connected to an array of column amplifiers. The signals required to achieve this are illustrated in Fig.~\ref{fig:star1000_row_read}:
\begin{figure}[hb!]
\begin{center}
\begin{tabular}{c} 
\includegraphics[height=6cm]{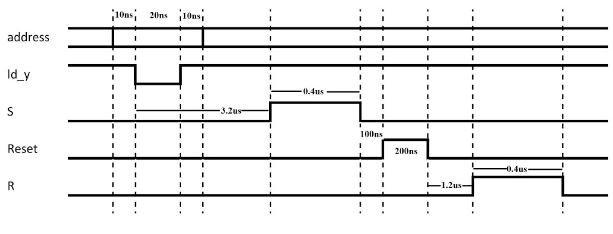}
\end{tabular}
\end{center}
\caption[]{ \label{fig:star1000_row_read} Timing diagram for row readout process.}
\end{figure} 
\begin{enumerate}
\item{Place proper $Y-$address on address bus.}
\item{Assert $Ld\verb|_|Y$ and latch the $Y-$address into the internal decoder circuitry.}
\item{Pulse $S$ signal which will sample values from all column pixels in the row into the column amplifier. This is the signal value after the integration time.} 
\item{Pulse `Reset' to reset the row in the pixel array. }
\item{Pulse $R$ signal which will sample the reset values of all column pixels in the row into the column amplifier. Eventually the output amplifier takes the difference between the reset level and the signal level and sends the analog output.}
\end{enumerate}
}

\item{`Cal' pulse to initialize the output amplifier: For every frame there at the first row readout, the `Cal' signal should be pulsed (illustrated in Fig.~\ref{fig:star1000_cal_pulse}). This gives the black reference (defined by the analog level at `Blackref' pin) value as output and thus calibrates the complete frame readout. The position of `Cal' pulse is with respect to the $S$ pulse and can be checked in the datasheet.}

\begin{figure}[h!]
\begin{center}
\begin{tabular}{c} 
\includegraphics[height=4cm]{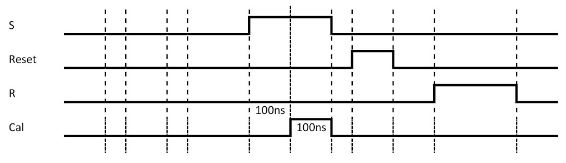}
\end{tabular}
\end{center}
\caption[]{ \label{fig:star1000_cal_pulse} Timing diagram for Cal pulse in row readout process}
\end{figure} 

\item{Pixel Readout/Column readout sequence: 
This process connects each pixel sequentially to the output programmable gain amplifier and gives reset pulses for the ADC to start conversion. The timing diagram for this process is illustrated in Fig.~\ref{fig:star1000_pixel_read}.}
\end{itemize}

\section{REAL TIME IMAGE PROCESSING}
\label{sec:processing}

The microblaze implemented on the FPGA can achieve a maximum performance of approximately 950 DMIPS, when operating at 100 MHz. This is sufficient to implement various image processing tasks for 1 Mp images captured by the sensors. Image processing tasks required in \emph{StarSense} are explained in this section. 

A star sensor is a wide-FOV camera with online image processing capabilities which reduces the image data to an attitude quaternion that describes the rotation of the sensor coordinate system with respect to the Earth-centered inertial (ECI) coordinate system. This is achieved by processing the images using multiple algorithms in sequence, implemented on the microblaze microprocessor.
All stars are point sources for the star sensor optics. The PSF of the optics is spread over $4\times4$ pixels on the detector. To pinpoint the location of a star, we use centroiding algorithm to calculate the position of the star images on the detector plane. Geometric voting algorithm compares the geometric patterns of the star centroids in the image with geometric patterns of stars in a catalog. In our case, the algorithm compares angles between all the pairs of stars in image with those in the catalog. A binary search method is used to find matches. After identifying the imaged stars, the quaternion of the rotation between the sensor coordinate system and ECI coordinate system is calculated by QUEST algorithm. These algorithms and their estimated performance are described in detail in a forthcoming paper (Ref.~\citenum{algorithms_paper}).

\section{CONCLUSION AND FUTURE WORK}
\label{sec:conclusion}  

We have developed a generic (application-wise) FPGA board to be used as image processor board on any space-based image sensor. This board includes
a method to generate the clocks required to read the image data, and a real-time image processor system which can be used for image processing tasks of different levels. Our immediate application is a UV telescope placed on the Moon (LUCI) and a star sensor \emph{StarSense} for minisatellites such as e.g. CubeSats.

The interface of this FPGA board with the star sensor \emph{StarSense} is complete, and calibration and testing is in progress. We have also completed the interface with the UV CCD on LUCI, and currently we are working on high-level data processing tasks implementation, such as, for example, obtaining light curves from time series data of astronomical objects and storing them on-board. 

In future, we are planning to use this FPGA board to operate an MCP-based detector employed on astronomical instruments that are currently in development (UV telescope and UV spectrograph, see Ref.~\citenum{balloon_paper_spie}). 

\acknowledgements
Part of this research has been supported by the Department of Science and Technology (Government of India) under Grant IR/S2/PU-006/2012.

\bibliography{report}

\begin{thebibliography}{1}

\bibitem{luci1}
M.~Safonova, J.~Mathew, and R.~Mohan, ``Prospect for uv observations from the
  moon,'' {\em ApSS} {\bf 353}, 329  (2014).

\bibitem{luci2}
J.~Mathew, A.~Prakash, M.~Sarpotdar, {\em et~al.}, ``Ultraviolet cosmic imager
  to study bright uv sources,'' {\em SPIE Astronomical Telescopes +
  Instrumentation, Space Telescopes and Instrumentation} {\bf 9905-146}
  (2016).

\bibitem{starsense1}
M.~Sarpotdar, J.~Mathew, A.~Prakash, {\em et~al.}, ``Design and development of
  a stars sensor - starsense,'' {\em ASI Meeting Srinagar}   (2016).

\bibitem{starsense2}
M.~Sarpotdar, J.~Mathew, A.~Sreejith, {\em et~al.}, ``Design and development of
  a star sensor-cum-asteroid tracker,'' {\em SPIE Astronomical Telescopes +
  Instrumentation, Space Telescopes and Instrumentation}   (2016).

\bibitem{datasheet}
O.~Semiconductors, ``Datasheet star1000 1 megapixel radiation hard cmos image
  sensor,'' {\em ON Semiconductors, Phoenix, Arizona, USA}   (January 2015).

\bibitem{algorithms_paper}
M.~Sarpotdar, J.~Mathew, A.~Sreejith, {\em et~al.}, ``Performance estimate of
  star sensor algorithms,'' {\em submitted to Experimental Astronomy}   (2016).

\bibitem{balloon_paper_spie}
A.~Sreejith, J.~Mathew, M.~Sarpotdar, {\em et~al.}, ``Balloon uv experiments
  for astronomical and atmospheric observations,'' {\em SPIE Astronomical
  Telescopes + Instrumentation}   (2016).

\end{thebibliography}
\bibliographystyle{spiejour}

\end{document}